\documentclass[aps,pra,twocolumn,superscriptaddress,showpacs]{revtex4-1}

\usepackage{graphicx}
\usepackage{amsmath}
\usepackage{color}

\usepackage[bookmarks=true,colorlinks=true,urlcolor=blue,linkcolor=blue,citecolor=blue,breaklinks]{hyperref}

\usepackage{cleveref}

\usepackage{amssymb}
\usepackage{booktabs}

\begin{document}

\sloppy

\title{First-principles Study of Ultrafast Dynamics of Dirac Plasmon in Graphene}

\newcommand*{\DIPC}[0]{{
Donostia International Physics Center (DIPC),
Paseo Manuel de Lardizabal 4, 20018 Donostia-San Sebasti\'an, Spain}}

\newcommand*{\IFS}[0]{{
Institute of Physics, Bijeni\v{c}ka 46, 10000 Zagreb, Croatia}}

\author{Dino Novko}
\email{dino.novko@gmail.com}
\affiliation{\IFS}
\affiliation{\DIPC}


\begin{abstract}
Exploring low-loss two-dimensional plasmon modes is considered central for achieving light manipulation at the nanoscale and applications in plasmonic science and technology. In this context, pump-probe spectroscopy is a powerful tool for investigating these collective modes and the corresponding energy transfer processes. Here, I present a first-principles study on non-equilibrium Dirac plasmon in graphene, wherein damping channels under ultrafast conditions are still not fully explored. The laser-induced blueshift of plasmon energy is explained in terms of thermal increase of the electron-hole pair concentration in the intraband channel. Interestingly, while damping pathways of the equilibrium graphene plasmon are entirely ruled by scatterings with acoustic phonons, the photoinduced plasmon predominantly transfers its energy to the strongly coupled hot optical phonons, which explains the experimentally-observed tenfold increase of the plasmon linewidth. The present study paves the way for an in-depth theoretical comprehension of plasmon temporal dynamics in novel two-dimensional systems and heterostructures.

\end{abstract}


\maketitle

\section{Introduction}

Understanding, and thus mastering, temporal dynamics of charge carriers in graphene and related quasi-two-dimensional materials is pivotal, but highly challenging task in material science. Many recent studies were devoted to explore the time evolution of the laser-excited electrons by means of time-resolved photoemission\,\cite{bib:johannsen13,bib:gierz13,bib:stange15,bib:tan17,bib:rohde18} and pump-probe optical absorption spectroscopies\,\cite{bib:kampfrath05,bib:sun08,bib:winnerl11,bib:jnawali13,bib:jensen14,bib:frenzel14,bib:mics15,bib:tomadin18} in graphene and graphite in order to reach the aforesaid goal. Precise time scales of ultrafast electron interactions were extracted, in particular, electron-electron scattering was shown to rule the dynamics below, while the coupling with the optical phonons (OP) above $\sim50$\,fs\,\cite{bib:kampfrath05,bib:johannsen13,bib:stange15,bib:tan17,bib:rohde18}. However, underlying microscopic processes still remain largely unexplored, mostly due to a lack of accompanying first-principles methodology that can quantitatively capture these features.

Photoinduced plasmon excitation, i.e., collective electron oscillations under highly non-equilibrium condition,\,\cite{bib:macdonald08,bib:huber16} is one such ultrafast phenomena that requires further insights. In graphene and graphene-based heterostructures, two-dimensional plasmons show quite exceptional features, e.g., electrical tunability\,\cite{bib:ju11,bib:fei12} and low losses\,\cite{bib:yan13,bib:woessner15,bib:ni18}, making these materials promising building blocks for optoelectronic and plasmonic devices. Recently, the relaxation dynamics of laser-induced graphene plasmon was monitored with unprecedented temporal and spatial resolution by using antenna-based near-field nanoscopy\,\cite{bib:wagner14,bib:ni16}. High electron temperatures achieved in these experiments are increasing the energy of graphene plasmon while concurrently increasing (decreasing) its linewidth (lifetime)\,\cite{bib:ni16}. The former was explained in terms of increase of the Drude weight, or equivalently increase of thermally excited electron-hole pairs, with elevated electron temperature, while the origin of the latter remains unresolved. Since the relaxation of equilibrium graphene plasmon was shown to be governed by the electron-phonon coupling\,\cite{bib:principi14,bib:woessner15,bib:ni18,bib:novko17}, mainly coupling with graphene acoustic phonons (AP)\,\cite{bib:principi14,bib:woessner15,bib:ni18}, it was speculated that the enhanced plasmon decay under non-equilibrium condition has the same origin\,\cite{bib:ni16}. However, projecting conclusions from the equilibrium situation might be premature, considering highly disparate thermal conditions in the two cases, but also having in mind the results extracted from time-resolved photoemission and optical absorption experiments where OP were proven to play a key role in relaxation processes\,\cite{bib:kampfrath05,bib:johannsen13,bib:stange15}.
Further quantitative analyses were thus far not provided, leaving us with many open questions regarding the ultrafast plasmon dynamics in graphene.

Here, I investigate the dynamics of laser-excited plasmon in lightly-doped graphene under non-equilibrium conditions by means of the robust \emph{ab initio} methodology. The work conjoins the electron-phonon coupling theory and the Coulomb screening in random phase approximation to capture the temperature-dependent plasmon decay due to phonons\,\cite{bib:shulga91,bib:novko16,bib:novko17}, while the non-equilibrium electron and phonon temperatures are simulated within the effective temperature model\,\cite{bib:allen87,bib:lin08,bib:perfetti07} with \emph{ab initio} input parameters\,\cite{bib:novko19b,bib:caruso19,bib:novko20}. I show, in agreement with previous reports\,\cite{bib:principi14,bib:woessner15,bib:ni18}, that graphene plasmon under equilibrium conditions (i.e., when electrons and phonons are thermalized) is predominantly decaying due to scattering with the AP. In particular, the obtained temperature dependence of the plasmon decay rate due to coupling with AP shows very good agreement with recent measurements done on high-mobility graphene\,\cite{bib:ni18}. However, the situation is drastically different for non-equilibrium conditions, where the majority of the laser-induced excess electron energy is transferred to the strongly coupled OP, creating hot phonon bath\,\cite{bib:kampfrath05}. In this case, the results show that the low-energy plasmons (i.e., at $\sim 0.1$\,eV), usually explored in the experiments\,\cite{bib:wagner14,bib:ni16,bib:ni18}, are mostly coupled to the hot OP, which is in contrast to the current belief\,\cite{bib:ni16}. The latter interaction is consequently responsible for the large time-dependent modifications of plasmon broadening. In addition, the laser excitation increases the phase space for the interband transitions (i.e., Landau damping), which in turn enhance the plasmon decay rate at higher energies (i.e., around $0.2$\,eV). Finally, the experimentally-observed ultrafast blueshift of plasmon energy is shown to be induced by the transient increase of the electron-hole pair exctations in the intraband channel (i.e., increase of the Drude weight). All in all, I believe that the \emph{ab initio} methodology and conclusions outlined here will be useful not only for comprehending the interplay of plasmon and phonon dynamics in graphene, but as well in other similar systems hosting a two-dimensional plasmon\,\cite{bib:sim15,bib:in18}.

%
%

\section{Theoretical methods}

\subsection{Three temperature model with \emph{ab initio} input parameters}

In order to simulate the laser-induced electron dynamics (i.e., electron-phonon thermalization) the three temperature model\,\cite{bib:perfetti07} with \emph{ab initio} input parameters is utilized (see also Refs.\,\cite{bib:novko19b,bib:caruso19}). 
Note that the recent time-resolved photoemission experiments\,\cite{bib:johannsen13,bib:tan17,bib:rohde18} have demonstrated that the nascent electron distribution is formed into Fermi-Dirac distribution almost instantly, i.e., within $25-50$\,fs, which justifies the use of the effective electron temperatures for the pump-probe time delays larger than 50\,fs.
In general, for a more quantitative description of electron dynamics below $50-100$\,fs one would need to adopt a more rigorous approaches, such as the classical\,\cite{bib:baranov14,bib:sadasivam17,bib:maldonado17} and quantum\,\cite{bib:schilp94} kinetic theories or the Keldysh Green's function technique\,\cite{bib:sentef13,bib:abdurazakov18}. Nevertheless, in graphene and graphite, due to mentioned quasi-instant electron thermalization as well as rapid electron-phonon coupling\,\cite{bib:hale11} both electron\,\cite{bib:ishida11,bib:johannsen13,bib:gierz13} and phonon\,\cite{bib:ishioka08,bib:yan09,bib:chatelain14} temperatures are well defined quantities already at the early stage of photoinduced electron dynamics.
Within the present model, the temperature of graphene is divided among three subsystems, i.e., electron temperature ($T_e$), temperature of the strongly coupled OP ($T_{\rm OP}$), and the remnant temperature that mostly belongs to AP ($T_{\rm AP}$). The energy flow between electron and phonon degrees of freedom is then dictated by the electron-phonon coupling\,\cite{bib:allen87}, while the thermalization between two phonon subsystems goes via anharmonic coupling. 
Since the quasi-instant formation of the Fermi-Dirac electron distribution is assumed, electron-hole recombination processes are not considered and only single (electron) temperature represents the dynamics of the electronic subsystem.

The out-of-equilibrium dynamics of the three subsystems can then be simulated by the following coupled equations:
\begin{eqnarray}
\frac{\partial  T_{e}}{\partial t} &=& \frac{I(t)}{\beta C_{e}}- \frac{G_{\rm OP}}{C_{e}}(T_{e}-T_{\rm OP}) - \frac{G_{\rm AP}}{C_{e}}(T_{e}-T_{\rm AP}),
\label{eq:eq1}
\\
\frac{\partial  T_{\rm OP}}{\partial t} &=& \frac{G_{\rm OP}}{C_{\rm OP}}(T_{e}-T_{\rm OP}) - \frac{T_{\rm OP}-T_{\rm AP}}{\tau},
\label{eq:eq2}
\\
\frac{\partial  T_{\rm AP}}{\partial t} &=& \frac{G_{\rm AP}}{C_{\rm AP}}(T_{e}-T_{\rm AP}) + \frac{C_{\rm OP}}{C_{\rm AP}}\frac{T_{\rm OP}-T_{\rm AP}}{\tau}.
\label{eq:eq3}
\end{eqnarray}
The corresponding specific heats 
$C_{e}$, $C_{\rm OP}$, and $C_{\rm AP}$ are 
defined as: 
\begin{eqnarray}
C_{e}
&=&
\int_{-\infty}^{\infty}
d\varepsilon N(\varepsilon)\varepsilon
\frac{\partial f(\varepsilon;T_{e})}{\partial T_{e}},
\label{eq:eq4}
\\
C_{\rm OP}
&=&
\int_{0}^{\infty}
d\omega F_{\rm OP}(\omega)\omega
\frac{\partial n(\omega;T_{\rm OP})}{\partial T_{\rm OP}},
\label{eq:eq5}
\\
C_{\rm AP}
&=&
\int_{0}^{\infty}
d\omega F_{\rm AP}(\omega)\omega
\frac{\partial n(\omega;T_{\rm AP})}{\partial T_{\rm AP}},
\label{eq:eq6}
\end{eqnarray}
where $N(\varepsilon)$ is electron, while $F_{\rm OP}(\omega)$ and $F_{\rm AP}(\omega)$ are phonon density of states. The electron-phonon relaxation rates $G_{\nu}$ are obtained from 
electron-phonon calculations via\,\cite{bib:lin08,bib:novko19b,bib:caruso19}: 
\begin{align}
G_{\nu}=\frac{\pi k_B}{\hbar N(\varepsilon_F)}\lambda_{\nu}\left\langle \omega^2 \right\rangle_{\nu}\int_{-\infty}^{\infty} d\varepsilon N^2(\varepsilon)\left(- \frac{\partial f(\varepsilon;T_{e})}{\partial \varepsilon} \right),
\label{eq:eq7}
\end{align}
where $\lambda_{\nu}$ are the electron-phonon coupling strengths for
$\nu=\mathrm{OP}$ and $\nu=\mathrm{AP}$,
\begin{align}
\lambda_{\nu}=\sum_{{\bf q}\mu}\lambda_{{\bf q}\mu}=2\int d\Omega \frac{\alpha^2 F_{\nu} (\Omega)}{\Omega},
\label{eq:eq7_0}
\end{align}
and $\left\langle \omega^2 \right\rangle_{\nu}$
are the corresponding second moments of the phonon spectrum, 
\begin{align}
\left\langle \omega^2 \right\rangle_{\nu}=\frac{2}{\lambda_{\nu}}\int d\Omega \Omega\alpha^2 F_{\nu} (\Omega).
\label{eq:eq7_1}
\end{align}
Here $\alpha^2 F(\Omega)$ is the Eliashberg function, which quantifies the amplitude of electron-phonon coupling for each phonon energy $\Omega$. The separation between the strongly coupled OP and weakly coupled AP are defined by introducing a cutoff $\lambda_c$ for the mode-resolved electron-phonon coupling strength $\lambda_{{\bf q}\mu}$. Namely, the modes that satisfy $\lambda_{{\bf q}\mu}<\lambda_c$ belong to the weakly-coupled, while the modes with $\lambda_{{\bf q}\mu}>\lambda_c$ to the strongly-coupled subsystem. For the clean separation $\lambda_c=1$ is used.

Furthermore, ${\tau}=3.3$\,ps is the anharmonic scattering time between the OP and AP modes\,\cite{bib:bonini07}. In addition, $I(t)$ describes a femtosecond Gaussian pump pulse with fluence $F$ and duration $t_p$,
\begin{align}
I(t)=\frac{2F}{t_p}\sqrt{\frac{\log{2}}{\pi}}\exp{\left[ -4\log{2}\left( \frac{t}{t_p} \right)^2 \right]}
\label{eq:eq7_2}
\end{align}
while $\beta$ determines the energy density of the pulse\,\cite{bib:johannsen13,bib:caruso19} and accounts for the fact that only a fraction of 
the pump fluence is absorbed by the sample. Here $\beta=400$, which gives very good agreement
with the experiment.

All the input parameters (except $\beta$) required in Eqs.~(\ref{eq:eq1}-\ref{eq:eq3}) have been determined entirely from first principles, i.e., by using density-functional theory and density-functional perturbation theory\,\cite{bib:baroni01}.

The shortcomings of the present effective temperature model are that the pump-pulse term $I(t)$ does not depend on laser frequency, but only on the laser power, as well as that it cannot describe the early electron dynamics before the formation of the Fermi-Dirac distribution. However, recent experiments have clearly demonstrated that the electron thermalization due to electron-electron interactions is almost instant (i.e., it happens within the first $25-50$\,fs)\,\cite{bib:johannsen13,bib:tan17,bib:rohde18}, and thus the discrete vertical electron excitations excited with the photon energy are almost instantly distributed into the hot Fermi-Dirac distribution, which can be described with the single electron temperature $T_e$. Therefore, the present model should be able to describe well the electron dynamics above 50\,fs and for the small excitation energies.

\subsection{Time-dependent electron excitation spectrum with phonon-assisted processes included}

In order to accurately describe electron excitation processes and optical absorption in doped single-layer graphene, the present paper employs a first-principles current-current formalism within linear response\,\cite{bib:novko16,bib:novko17}. Within this approach the bare current-current correlation function $\pi_{\alpha\alpha}$ (where $\alpha$ is the polarization direction) is screened with Coulomb interaction by the following Dyson equation $\widetilde{\pi}_{\alpha\alpha}=\pi_{\alpha\alpha}+\pi_{\alpha\alpha}\otimes D\otimes \widetilde{\pi}_{\alpha\alpha}$ (where $D$ is the photon propagator). The corresponding optical conductivity at the given frequency $\omega$ is $\sigma_{\alpha\alpha}(\omega)=i\pi_{\alpha\alpha}(\omega)/\omega$, while the excitation spectrum that includes collective modes is given with $S(\mathbf{q},\omega)\propto {\rm Im}\,\widetilde{\pi}_{\alpha\alpha}(\mathbf{q},\omega)/\omega$\,\cite{bib:novko16,bib:novko17}. To include electron-phonon coupling into excitation spectrum as well as plasmon-phonon coupling the current-current correlation function is first separated into interband and intraband contributions and then the electron-phonon coupling is incorporated into the latter channel\,\cite{bib:novko17}.

The interband term has the following form
\begin{eqnarray}
\pi^{\mathrm{inter}}_{\alpha\alpha}(\omega; T_e)&=&\frac{2}{A}\sum_{\mathbf{k},n\neq m}\frac{\omega\left|j^{\alpha}_{nm\mathbf{k}}\right|^2}{\varepsilon_{m\mathbf{k}}-\varepsilon_{n\mathbf{k}}}\nonumber\\
&&\times\frac{f(\varepsilon_{n\mathbf{k}}; T_e)-f(\varepsilon_{m\mathbf{k}}; T_e)}{\omega+\varepsilon_{n\mathbf{k}}-\varepsilon_{m\mathbf{k}}+i\gamma_{\mathrm{inter}}},
\label{eq:eq8}
\end{eqnarray}
where $\varepsilon_{n\mathbf{k}}$ is the electron energy for band index $n$ and electron momentum $\mathbf{k}$, $f(\varepsilon_{n\mathbf{k}};T_e)$ is the Fermi-Dirac distribution function at electron temperature $T_e$, $\gamma_{\rm inter}$ is the phenomenological relaxation parameter for interband transitions\,\cite{bib:novko16}, and $A$ is the area of the unit cell. The current vertex (i.e., electron-photon coupling function) is defined as,
\begin{eqnarray}
j^{\alpha}_{nm\mathbf{k}}=\frac{\hbar e}{2i m}\int_{\Omega} d{\bf r} \left\{\phi_{n{\bf k}}^*({\bf r})\partial_\alpha\phi_{m{\bf k}}({\bf r})\right.
\nonumber\\
\nonumber\\
\left.-[\partial_\alpha\phi_{n{\bf k}}^*({\bf r})]\phi_{m{\bf k}}({\bf r})\right\},
\label{eq:eq8_0}
\end{eqnarray}
where $\phi_{n{\bf k}}$ are the Kohn-Sham ground state wavefunctions.

The intraband current-current correlation function with electron-phonon interaction included can be written as\,\cite{bib:allen71,bib:novko17}
\begin{eqnarray}
\label{eq:eq9}
\pi^{\mathrm{intra}}_{\alpha\alpha}(\omega;\{T\})&&=\frac{2}{\Omega}\sum_{\mathbf{k},n}\left[-\frac{\partial f(\varepsilon_{n\mathbf{k}};T_e)}{\partial\varepsilon_{n\mathbf{k}}}\right]\left|j^{\alpha}_{nn\mathbf{K}}\right|^2\nonumber\\
&&\times
\frac{\omega}{\omega\left[1+\lambda_{\mathrm{ep}}(\omega;\{T\})\right]+i\gamma_{\mathrm{ep}}(\omega;\{T\})}.
\end{eqnarray}
I have obtained this result by applying the Holstein theory for normal metals, where the conductivity is calculated by means of a  diagrammatic  analysis  and  solving  the  Bethe-Salpeter  equation for the electron-phonon interaction\,\cite{bib:holstein64,bib:allen74,bib:kupcic15}.
Here the electron-phonon coupling is incorporated through dynamical electron-hole pair energy renormalization and decay rate parameters, i.e., $\lambda_{\rm ep}(\omega;\{T\})$ and $\gamma_{\rm ep}(\omega;\{T\})$, both of which are dependent on electron and phonon temperatures, i.e., $T_e$ and $T_{\rm ph}$\,\cite{bib:allen74,bib:shulga91,bib:novko17}. In particular, dynamical electron-hole pair decay rate due to electron-phonon coupling is\,\cite{bib:shulga91,bib:novko17,bib:novko20}
\begin{eqnarray}
\gamma_{\rm ep}(\omega;\{T\})
=
\frac{\pi}{\omega}
\int d\Omega \alpha^2 F(\Omega)
\left[
2\omega\coth\frac{\Omega}{2k_BT_{\rm ph}}\right. \nonumber\\ 
\left.-
(\omega+\Omega)\coth\frac{\omega+\Omega}{2k_BT_{e}}
+
(\omega-\Omega)\coth\frac{\omega-\Omega}{2k_BT_{e}}
\right].
\label{eq:eq10}
\end{eqnarray}
The energy renormalization parameter $\lambda_{\rm ep}(\omega;\{T\})$ is obtained by Kramers-Kronig transformation of $\gamma_{\rm ep}(\omega;\{T\})$. By dividing the electron and phonon degrees of freedom into three subsystems defined with $T_e$, $T_{\rm OP}$, and $T_{\rm AP}$, as in the previous subsection, one can separate $\gamma_{\rm ep}(\omega;\{T\})$ in the three parts\,\cite{bib:novko20}
\begin{eqnarray}
\gamma_{\rm ep}(\omega;\{T\})
&=&
\gamma_{\rm ep}(\omega;T_{e})
+
\gamma_{\rm ep}(T_{\rm OP})
+
\gamma_{\rm ep}(T_{\rm AP})
,
\end{eqnarray}
where
\begin{align}
&\gamma_{\rm ep}(\omega;T_{e})
=
-
\frac{\pi}{\omega}
\int d\Omega \alpha^2 F(\Omega)
\left[
(\omega+\Omega)\coth\frac{\omega+\Omega}{2k_BT_{e}}\right.\nonumber\\
&\left.-
(\omega-\Omega)\coth\frac{\omega-\Omega}{2k_BT_{e}}
\right],
\label{eq:eq11}
\\
&\gamma_{\rm ep}(T_{\rm OP})
=
2\pi
\int d\Omega \alpha^2 F_{\rm OP}(\Omega)
\coth\frac{\Omega}{2k_BT_{\rm OP}}
,
\\
&\gamma_{\rm ep}(T_{\rm AP})
=
2\pi
\int d\Omega \alpha^2 F_{\rm AP}(\Omega)
\coth\frac{\Omega}{2k_BT_{\rm AP}}
.
\label{eq:eq12}
\end{align}
The Eliashberg function is accordingly also divided into contributions coming from the AP and OP phonon subsystems, i.e., $\alpha^2 F(\Omega)=\alpha^2 F_{\rm AP}(\Omega)+\alpha^2 F_{\rm OP}(\Omega)$.

Finally, the excitation spectrum with phonon-assisted decay channels and full temperature dependence included is calculated as
\begin{eqnarray}
S(\mathbf{q},\omega;\{T\})\propto \frac{{\rm Im}\,\widetilde{\pi}_{\alpha\alpha}(\mathbf{q},\omega;\{T\})}{\omega},
\label{eq:eq13}
\end{eqnarray}
where $\widetilde{\pi}_{\alpha\alpha}$ is the full (intraband and interband) current-current response tensor that includes Coulomb screening and electron-phonon interactions as presented above.

The time dynamics of the electron excitations and plasmons is then obtained by correlating Eq.\,\eqref{eq:eq13} with equations for the time evolution of the effective temperatures Eqs.\,\eqref{eq:eq1}-\eqref{eq:eq3}\,\cite{bib:novko19,bib:novko20}.

Note that the present methodology includes both intraband and interband electronic transitions and thus accounts for both low- and high-energy long-wavelength excitations. Therefore, it could be applied both for lightly- and heavily-doped samples, as well as to any kind of plasmonic materials such as metals, semimetals, or even gapped systems\,\cite{bib:scholz13,bib:iurov16}.

\subsection{Computational details}

The ground-state calculations were done by means of the {\sc quantum espresso} (QE) package\,\cite{bib:qe1} with a plane-wave cutoff energy of 50\,Ry. Norm-conserving pseudopotentials were used with the LDA exchange-correlation functional\,\cite{bib:lda}. A $24\times24\times1$ Monkhorst-Pack grid was used for sampling the Brillouin zone (with Gaussian smearing of 0.02\,Ry). Electron and hole dopings were simulated by adding and removing, respectively, the electrons and introducing the compensating homogeneous charged background. Phonon energies and electron-phonon matrix elements, which are needed for obtaining the input parameters for the three temperature model Eqs.\,\eqref{eq:eq1}-\eqref{eq:eq3} as well as for the decay rates due to electron-phonon coupling Eqs.\,\eqref{eq:eq10}-\eqref{eq:eq12}, are obtained by using density functional perturbation theory\,\cite{bib:baroni01} as implemented in QE. The Eliashberg function $\alpha^2 F(\Omega)$ is calculated on $400\times400\times1$ and $40\times40\times1$ electron and phonon momentum grids, respectively.
The electron momentum summations in the intraband and interband current-current correlation functions are done on a $400\times400\times1$ grid including up to 20 unoccupied electronic bands.

%
%

\section{Results and discussion}

%
\begin{figure}[!t]
\includegraphics[width=0.48\textwidth]{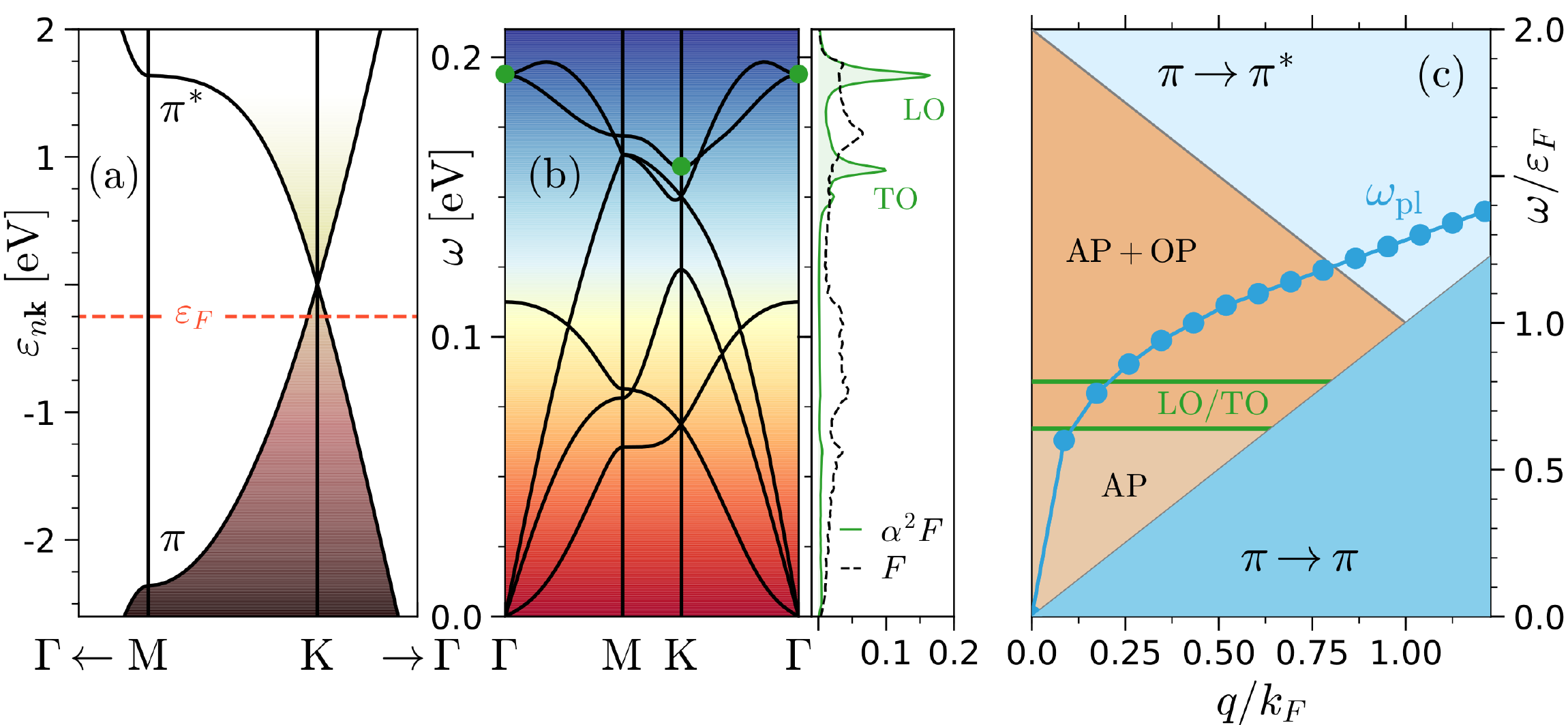}
\caption{\label{fig:fig1}(a) Electron and (b) phonon band structures of lightly hole-doped graphene. Dashed red line in (a) is the Fermi level, while the shaded brown area schematically represents the skewed electron distribution for finite electron temperature. Green and black dashed lines in panel (b) are Eliashberg function and phonon density of states, respectively. Red and blue colors depict the Bose-Einstein distribution of phonon modes at finite phonon temperature. (c) Plasmon dispersion in graphene for Fermi energy $\varepsilon_F=-250$\,meV (blue circles). Blue and light blue regions are the intraband and interband excitations regions, respectively. The phase space where plasmon couples only to acoustic (light orange) as well as both acoustic and optical (orange) phonon modes is shown as well.
}
\end{figure}
%

Ultrafast electron dynamics is explored here for the hole-doped graphene where the Fermi energy is $\varepsilon_F=-250$\,meV, as it is the case, e.g., for graphene adsorbed on SiC surface\,\cite{bib:johannsen13} (note that the conclusions of the paper would be the same for lightly electron-doped graphene as in the graphene/SiO$_2$ system\,\cite{bib:ni16} due to electron-hole symmetry in the low-energy region of band structure). The corresponding electron and phonon band structures are shown in Figs.\,\ref{fig:fig1}(a) and \ref{fig:fig1}(b). Additionally, Fig.\,\ref{fig:fig1}(b) shows the results for the Eliashberg function that measures the degree of the electron-phonon coupling with energy resolution. As is well know\,\cite{bib:piscanec04}, electrons are strongly coupled to OP at K and $\Gamma$ point of the Brillouin zone, i.e., $\omega\gtrsim 0.16$\,eV where Eliashberg function shows two prominent peaks, while only weakly coupled to the rest of the modes (mostly low-energy AP). Accordingly, the plasmon excitations below $0.16$\,eV are weakly coupled to AP\,\cite{bib:principi14}, while at higher energies plasmon decay is mostly due to the strong coupling with OP\,\cite{bib:novko17}. The graphene plasmon dispersion $\omega_{\rm pl}$ and the corresponding decay regions due to intraband and interband transitions (Landau damping)\,\cite{bib:wunsch06} are shown in Fig.\,\ref{fig:fig1}(c).

%
\begin{figure}[!t]
\includegraphics[width=0.44\textwidth]{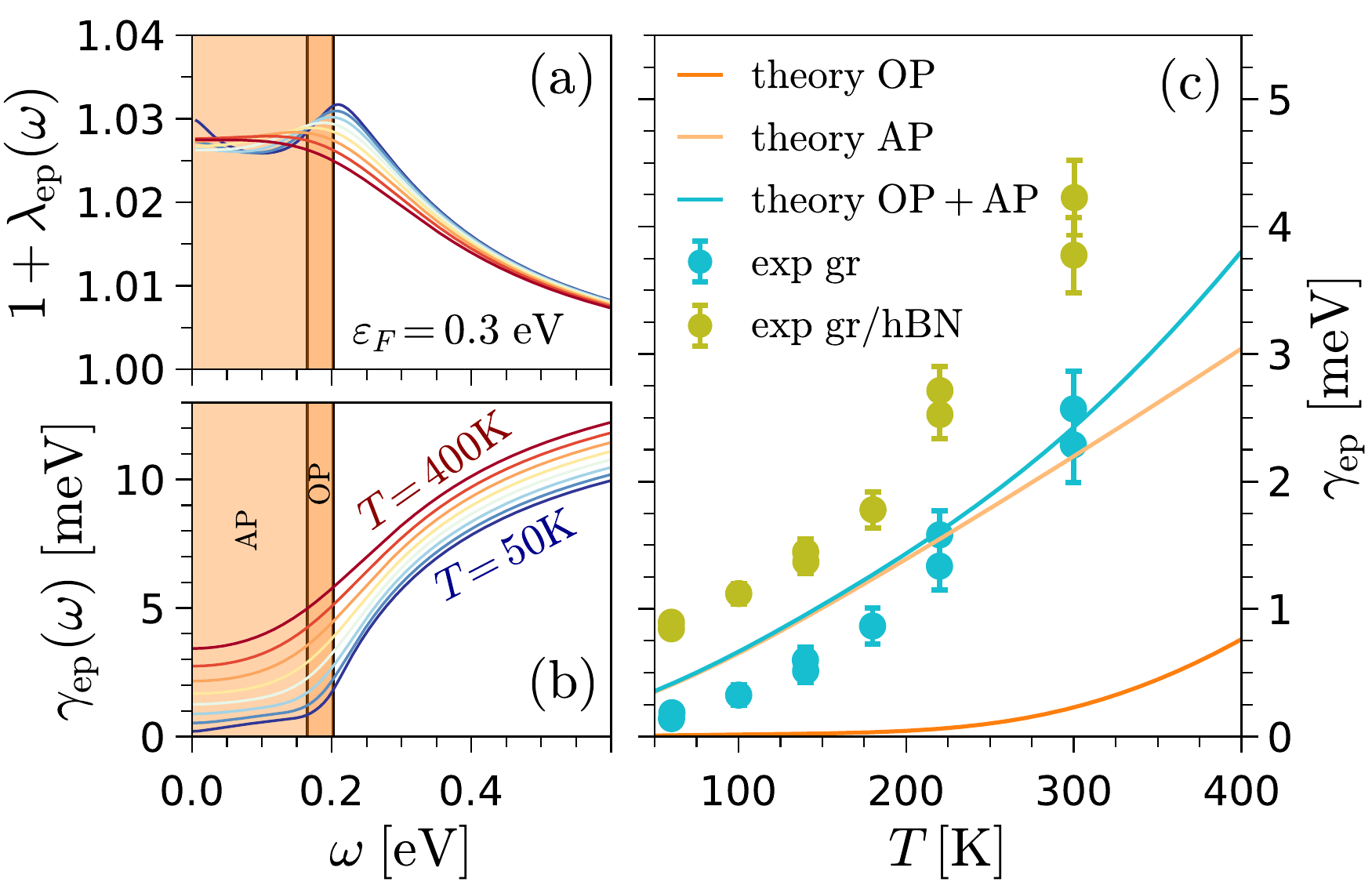}
\caption{\label{fig:fig2}(a) Energy renormalization $1+\lambda_{\rm ep}(\omega)$ and (b) decay rate $\gamma_{\rm ep}(\omega)$ of the electron-hole pairs due to coupling with phonon as a function of frequency and temperature. The energy windows of acoustic and optical phonons are highlighted with light orange and orange areas, respectively. (c) Temperature dependence of $\gamma_{\rm ep}(\omega_{\rm pl})$ when plasmon energy is $\omega_{\rm pl}=110$\,meV (blue line). The contributions coming from acoustic and optical phonons are shown with light orange and orange lines. The experimentally determined plasmon decay rate for graphene encapsulated in hexagonal boron nitride (green circles) is shown for comparison\,\cite{bib:ni18}. The corresponding decay rate without the influence of the boron nitride is shown with blue circles.
}
\end{figure}
%

Figures \ref{fig:fig2}(a) and \ref{fig:fig2}(b) further depict the energy renormalization $1+\lambda_{\rm ep}(\omega)$ and decay rate $\gamma_{\rm ep}(\omega)$, respectively, of the electron-hole pairs due to coupling with phonons as a function of temperature\,\cite{bib:allen71,bib:allen74,bib:shulga91,bib:kupcic15}. The Fermi energy is here chosen to be $\varepsilon_F=300$\,meV for the sake of comparison with the experiment\,\cite{bib:ni18} (the rest of the results are for $\varepsilon_F=-250$\,meV as mentioned earlier). For plasmon energies, i.e., $\omega=\omega_{\rm pl}$, the quantities $1+\lambda_{\rm ep}(\omega_{\rm pl})$ and $\gamma_{\rm ep}(\omega_{\rm pl})$ are equivalent to the plasmon energy renormalization and plasmon decay rate due to coupling with phonons\,\cite{bib:novko17}. The results show that the plasmon energy is insignificantly renormalized due to electron-phonon coupling, with very small temperature modifications. The corresponding plasmon broadening is as well small (especially for $\omega_{\rm pl}<0.16$\,eV, where electrons mostly couple to AP) but notable. For the temperature range presented here, the temperature-induced change in plasmon broadening is more pronounced for the lower energies since AP are more easily excited with temperature than the high-energy OP (i.e., $\omega_{\rm OP}\gg k_B T$). Figure \ref{fig:fig2}(c) shows the temperature dependence of the plasmon decay rate when $\omega_{\rm pl}= 110$\,meV. The results are in very good agreement with the decay rate extracted from the recent experiment done on high-mobility graphene\,\cite{bib:ni18}. Further analysis demonstrates that the decay rate of the equilibrium graphene plasmon and its temperature dependence predominantly comes from coupling with AP, in agreement with the previous studies\,\cite{bib:principi14,bib:woessner15,bib:ni18}.

%
\begin{figure}[!t]
\includegraphics[width=0.48\textwidth]{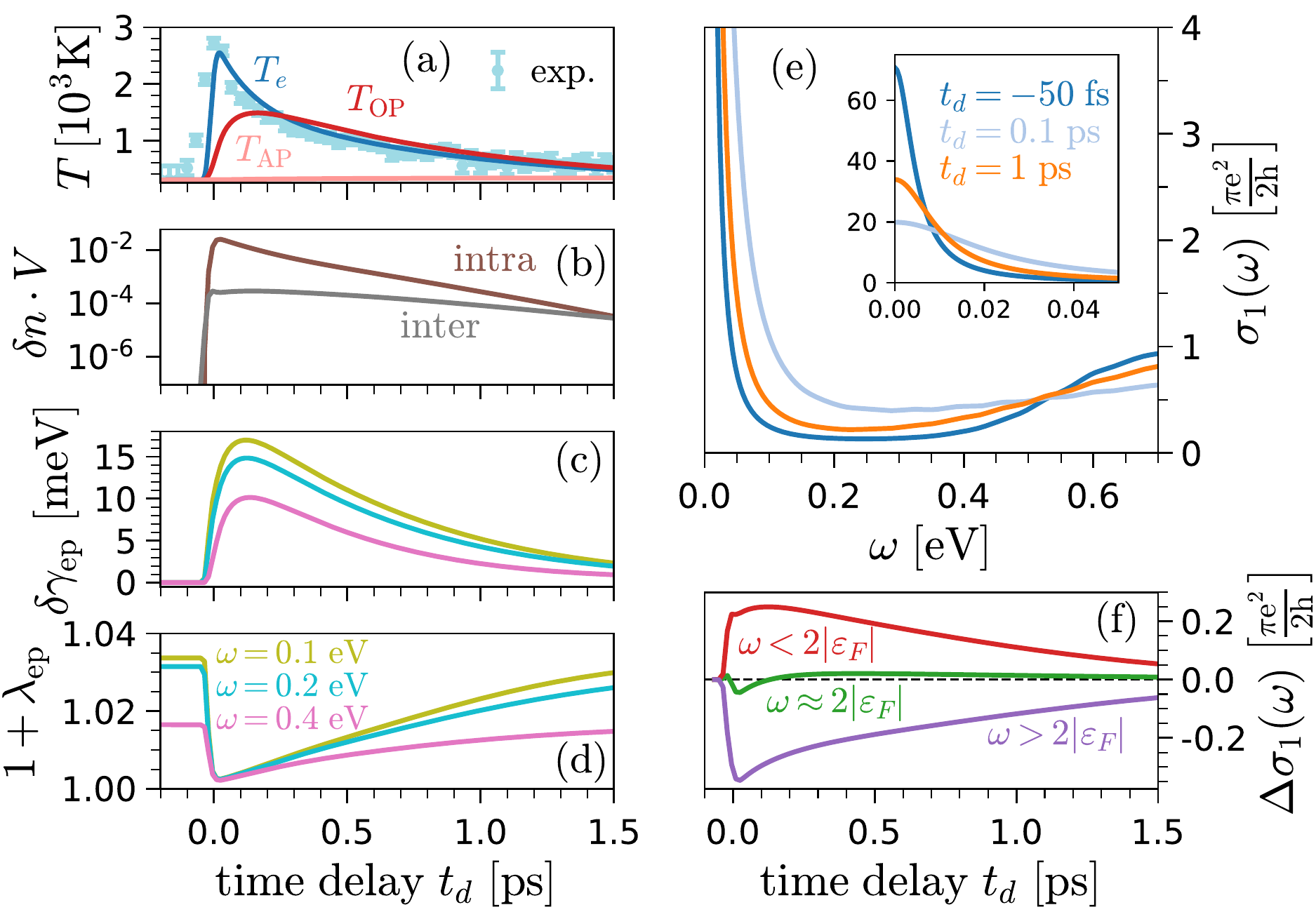}
\caption{\label{fig:fig3}(a) Time evolution of electron temperature $T_e$, temperature of the strongly coupled optical phonons $T_{\rm OP}$, and temperature of the remnant phonon, mostly acoustic, modes $T_{\rm AP}$. The laser with the fluence of $8$\,J/m$^2$ excites the system at the zero delay time. The extracted experimental results for $T_e$\,\cite{bib:gierz13} are shown with blue circles. (b) The corresponding photoinduced charge density modifications for intraband and interband channels. Time dependence of (c) electron-phonon decay rate changes $\delta\gamma_{\rm ep}$ and (d) energy renormalization parameter $1+\lambda_{\rm ep}$ for three different values of excitation energy $\omega$. (e) Transient optical absorption $\sigma_1(\omega)$ for three different values of time delay. Inset: The corresponding low-energy part (Drude peak). (f) Time evolution of photoconductivity $\Delta\sigma_1(\omega)$ for excitation energies below, around, and above the interband threshold $2|\varepsilon_F|$.
}
\end{figure}
%

I turn now to the study of the non-equilibrium condition, i.e., of the ultrafast electron dynamics in graphene by means of the three temperature model with \emph{ab initio} input parameters.
The resultant time evolution of these temperatures is shown in Fig.\,\ref{fig:fig3}(a), where the laser with fluence of $F=8$\,J/m$^2$ and duration of 30\,fs excites the system at time delay $t_d=0$. Right after the laser excitation, the electron temperature $T_e$ abruptly increases up to almost 3000\,K and subsequently decays due to coupling with OP, in good agreement with the experiment\,\cite{bib:gierz13}. Consequently, the temperature of the strongly coupled OP elevates above 1000\,K, while the AP remain almost at the same temperature. Since the energy exchange rate between phonon and electron baths is proportional to $\lambda\left\langle \omega^2 \right\rangle / C_{\rm ph}$ (where $\lambda$ is the electron-phonon coupling strength, $\left\langle \omega^2 \right\rangle$ is the second moment of the phonon spectrum, and $C_{\rm ph}$ is the heat capacity of the relevant phonon subsystem)\,\cite{bib:allen87}, the obtained dramatic difference between the OP and AP temperatures comes not only because the OP are coupled more strongly to the electrons than the AP ($\lambda_{\rm OP}\left\langle \omega^2 \right\rangle_{\rm OP}>\lambda_{\rm AP}\left\langle \omega^2 \right\rangle_{\rm AP}$), but also because the strongly coupled OP subsystem consist of only very few modes around $\Gamma$ and K points of the Brillouin zone [see Fig.\,\ref{fig:fig1}(b)] and thus $C_{\rm OP}\ll C_{\rm AP}$. Such laser-induced hot OP scenario was already discussed in various spectroscopy studies\,\cite{bib:kampfrath05,bib:johannsen13,bib:stange15}.

Figure \ref{fig:fig3}(b) shows the ensuing charge density modifications $\delta n$ as a function of time delay. As laser excites the system, electron-hole pair concentration increases both in intraband (electron and hole are in the same band) and interband (electron and hole are in different bands) channels. In fact, for the lightly-doped graphene with $\varepsilon_F=-250$\,meV it turns out that $\delta n_{\rm intra} \gg \delta n_{\rm inter}$. The laser-induced changes of the electron-phonon coupling are shown in Figs.\,\ref{fig:fig3}(c) and \ref{fig:fig3}(d). In particular, time-dependant modifications of the electron-phonon decay rate $\delta\gamma_{\rm ep}$ and energy renormalization parameter $1+\lambda_{\rm ep}$ are depicted for three different excitation energies $\omega$. Since $1+\lambda_{\rm ep}$ is already small for the equilibrium situation, it is not surprising that $1+\lambda_{\rm ep}$ experiences only minor changes as a function of pump-probe time delay and excitation energy. One can also note that laser excitation reduces the energy renormalization parameter as a function of time. On the other hand, for the same laser conditions, the photoinduced decay rate modifications $\delta\gamma_{\rm ep}$ are relatively high [i.e., $\delta\gamma_{\rm ep}\gg \gamma_{\rm ep}(T=300\,{\rm K})$] and actually follow the variations of both $T_e$ and $T_{\rm OP}$ (see the discussion below). Also, the overall intensity of $\delta\gamma_{\rm ep}$ over time is bigger for smaller excitation energies (i.e., for $\omega=0.1$\,eV and 0.2\,eV, compared to $\omega=0.4$\,eV). This is because for $\omega\lesssim 0.2$\,eV the equilibrium value of $\gamma_{\rm ep}$ is small and includes mostly the contributions from the weakly-coupled AP modes, while when both $T_e$ and $T_{\rm OP}$ are elevated the probability of scattering on the OP, which are strongly coupled to electrons, increases significantly. A more rapid increase of electron-OP scattering probability when $T\gtrsim 300$\,K can, for example, be seen in Fig.\,\ref{fig:fig2}(c). However, when $\omega\gg 0.2$\,eV the probability of scattering on the OP is less altered for the present laser conditions, since $\omega> T_e, T_{\rm OP}$. Therefore, the relative modification of damping rate is less pronounced for $\omega\gg 0.2$\,eV compared to $\omega\lesssim 0.2$\,eV.

%
\begin{figure}[!t]
\includegraphics[width=0.48\textwidth]{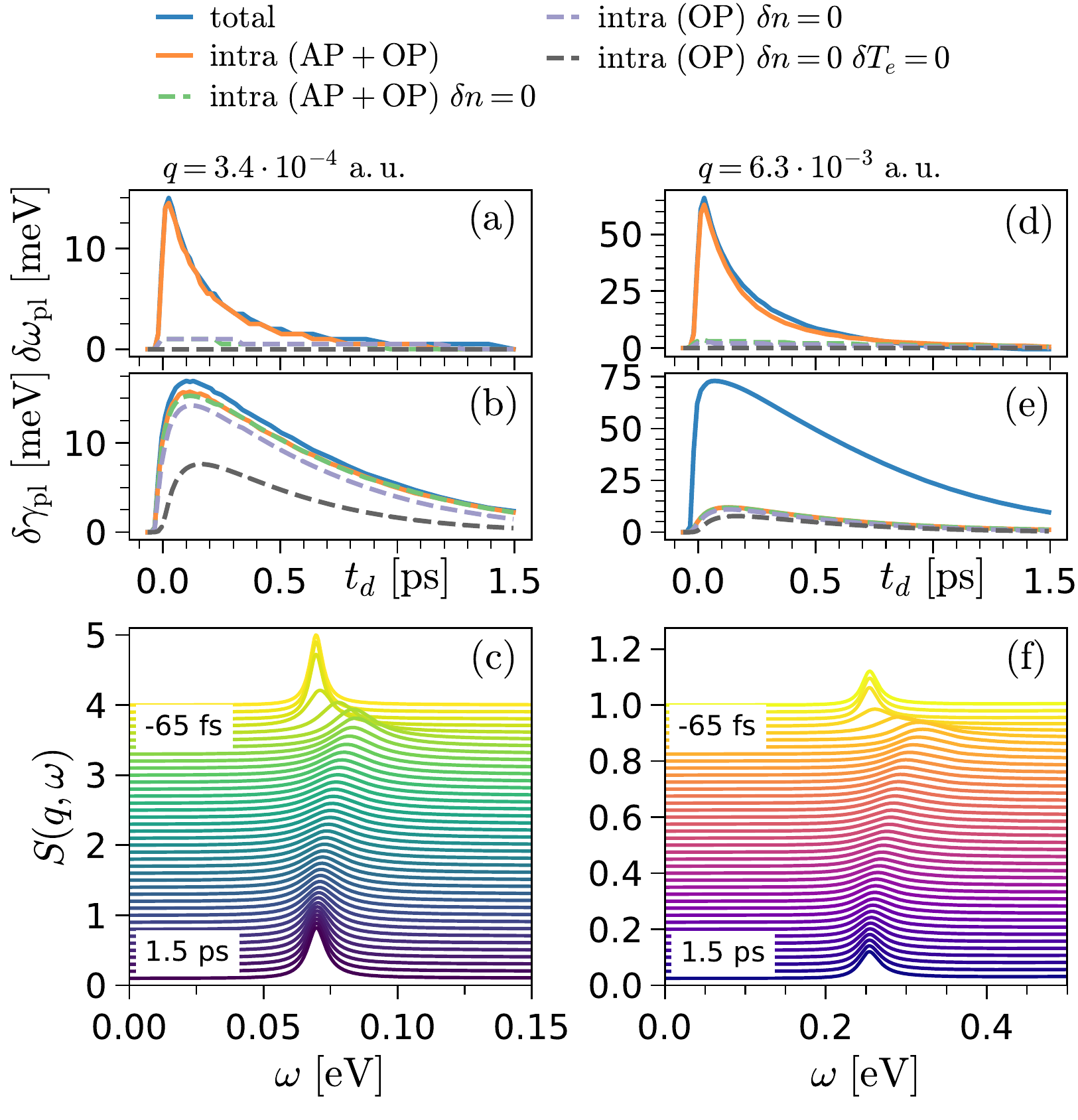}
\caption{\label{fig:fig4} Modifications of plasmon (a) energy $\delta\omega_{\rm pl}$ and (b) linewidth $\delta\gamma_{\rm pl}$ as well as (c) the corresponding spectral function $S(q,\omega)$ as a function of pump-probe time delay $t_d$ when the plasmon wavevector is $q=3.4\cdot10^{-4}$\,a.u. Different contributions to $\delta\omega_{\rm pl}$ and $\delta\gamma_{\rm pl}$ are shown: intraband + interband excitations (blue), intraband excitations with the full electron-phonon coupling (orange), intraband excitations without changes in Drude weight $\delta n=0$ (green), intraband excitations when $\delta n=0$ and without electron-acoustic phonon (AP) scattering (purple), and intraband excitations when $\delta n=0$, without AP, and without electron temperature $T_e$ contribution to electron-optical phonon (OP) scattering rate (black).
(d)-(f) Same as in (a)-(c) but for larger plasmon wavevector, i.e., $q=6.3\cdot10^{-3}$\,a.u.
Note the different spectral-intensity and energy scales in (c) and (f) panels}
\end{figure}
%

The time dynamics of charge density and decay rate are relevant for comprehending the transient optical absorption, i.e., photoconductivity, as well as the relaxation mechanisms of non-equilibrium plasmons. Figure \ref{fig:fig3}(e) depicts the time evolution of optical absorption $\sigma_1(\omega)$ (i.e., the real part of optical conductivity $\sigma$) up to 0.7\,eV. The observed modifications are due to photoinduced variations of $T_e$, and also due to changes in decay rate $\delta\gamma_{\rm ep}$ (note that $\sigma_1(0)\propto n/\gamma_{\rm ep}$ and $\sigma_1(\omega>0)\propto n\gamma_{\rm ep}/(\omega^2+\gamma_{\rm ep}^2)$\,\cite{bib:allen71,bib:kupcic15,bib:novko17}). In particular, the low-energy part of $\sigma_1(\omega)$ decreases significantly around $t_d=0$ due to changes in $\delta\gamma_{\rm ep}$ and then it starts to increase back to its equilibrium value [see the inset in Fig\,\ref{fig:fig3}(e)]. Contrary, for higher excitation energies up to around interband threshold $2|\varepsilon_F|$ the photoinduced modifications of $\sigma_1(\omega)$ are first increasing and then decreasing. The time evolution of photoconductivity $\Delta\sigma_1$ below, around, and above interband threshold $2|\varepsilon_F|$ is also depicted in Fig.\,\ref{fig:fig3}(f). In all these three energy regimes, the photoconductivity $\Delta\sigma_1$ shows different time behaviour, which is simply due to modifications of the interband onset when $T_e$ is altered.

Note that the experimentally observed negative and positive values of graphene photoconductivity over different values of $\omega$ are widely discussed and analyzed in literature\,\cite{bib:kampfrath05,bib:sun08,bib:winnerl11,bib:jnawali13,bib:jensen14,bib:frenzel14,bib:mics15,bib:tomadin18}. This shows that the present methodology could be also useful for studying the non-equilibrium properties in pump-probe optical spectroscopy and non-equilibrium transport via $\sigma_{\alpha\alpha}(\omega;\{T\})$ and $\sigma_{\alpha\alpha}(\omega=0;\{T\})$, respectively. However,
here the focus is more on the ultrafast plasmons and the corresponding dynamics in different energy regimes. The time modulations of graphene plasmon properties (e.g., energy loss and Drude weight) under optical pumping were in fact discussed recently\,\cite{bib:page16,bib:hamm16,bib:sun16,bib:petersen17,bib:wilson18}, however, the detailed \emph{ab-initio} study on the corresponding plasmon loss channels is still lacking.

Figures \ref{fig:fig4}(a)-(c) and \ref{fig:fig4}(d)-(f) show the variations in the plasmon energy $\delta\omega_{\rm pl}$, plasmon linewidth $\delta\gamma_{\rm pl}$, and the spectral function $S(q,\omega)$ as a function of time delay $t_d$ for two different values of wavevector $q$, i.e., for two different energy regimes (note that $\omega_{\rm pl}\propto \sqrt{q}$). Note that the time variations of the plasmon energy of 2D system can be calculated as\,\cite{bib:kupcic14}:
\begin{align}
\omega_{\rm pl}^2(\mathbf{q},\omega;\{T\}) = 2\pi q_{\alpha} \omega {\rm Im}\left[ \sigma_{\alpha\alpha}^{\rm intra}(\mathbf{q},\omega;\{T\})\right.\nonumber\\
\left.+\sigma_{\alpha\alpha}^{\rm inter}(\mathbf{q},\omega;\{T\}) \right],
\label{eq:plen}
\end{align}
while the corresponding time evolution of the plasmon linewidth as
\begin{align}
\gamma_{\rm pl}(\mathbf{q},\omega;\{T\}) = 2\pi q_{\alpha} \hbar {\rm Re}\left[ \sigma_{\alpha\alpha}^{\rm intra}(\mathbf{q},\omega;\{T\})\right.\nonumber\\
\left.+\sigma_{\alpha\alpha}^{\rm inter}(\mathbf{q},\omega;\{T\}) \right].
\label{eq:plen1}
\end{align}
Also note that damping due to electron-phonon coupling enters the intraband, while the Landau damping the interband part of time-dependent optical conductivity.

The obtained time dynamics of the photoinduced graphene plasmon is in good agreement with the experimental observations\,\cite{bib:wagner14,bib:ni16}. Namely, the results show that the graphene plasmon is blueshifted and broadened upon the laser excitation. Also, $\delta\omega_{\rm pl}$ and $\delta\gamma_{\rm pl}$ are notably more pronounced for larger plasmon energies $\omega_{\rm pl}$ (larger wavectors $q$). In order to understand these transient features of graphene plasmon, it is necessary to dissect different contributions to $\omega_{\rm pl}$ and $\gamma_{\rm pl}$. First of all, note that in general both intraband and interband excitations determine the total value of plasmon energy. Here, the laser-induced modifications of plasmon energy predominantly come from the increase of electron-hole pair concentrations in the intraband channel $n_{\rm intra}$ [see Figs.\,\ref{fig:fig4}(a) and \ref{fig:fig4}(d)]. In other words, laser elevates $T_e$, i.e., increases the number of thermally excited electrons in the intraband channel [see also Fig.\,\ref{fig:fig3}(b)], which in turn increases the Drude weight and thus the plasmon energy\,\cite{bib:wagner14,bib:ni16}. As already discussed above, the renormalization of the plasmon energy due to electron-phonon coupling is insignificant [see also Fig.\,\ref{fig:fig3}(d)].

The processes underlying the plasmon decay rate $\delta\gamma_{\rm pl}$ as a function of time delay are a bit more complex than the processes ruling $\delta\omega_{\rm pl}$. For plasmon energies $\omega_{\rm pl}\approx 0.1$\,eV, it turns out that the increase of the plasmon broadening $\delta\gamma_{\rm pl}$ is entirely ruled by electron-phonon coupling. What is intriguing and actually at odds with the previous assumptions\,\cite{bib:ni16}, is that the photoinduced plasmon decays mostly due to scatterings with the OP modes [cf. green and purple dashed lines in Fig.\,\ref{fig:fig4}(b)]. This is also in contrast with the decay mechanisms of the equilibrium plasmon, for which the scattering with the AP modes is the main loss channel [see Fig.\,\ref{fig:fig2}(c)]. The present analysis also shows that part of the electron-OP scattering contribution to the $\delta\gamma_{\rm pl}$ is induced by the elevated $T_e$ and part by the elevated $T_{\rm OP}$ [cf. purple and black dashed lines in Fig.\,\ref{fig:fig4}(b)]. Namely, the laser heats the electrons (i.e., elevates $T_e$), which in turn increases the electron phase space for the electron-OP scattering, especially when $T_e > \omega_{\rm OP}$. In addition, hot electrons transfer the excess energy to the strongly-coupled OP, i.e., $T_{\rm OP}$ rises substantially, which increases the number of thermally excited OP and therefore increases $\delta\gamma_{\rm pl}$. All in all, the results show that the plasmon-OP scattering is behind the experimentally-observed plasmon broadening under non-equilibrium conditions. The coupling with the OP modes is generally much more stronger than with the AP modes. However, for equilibrium case the energy conservation condition $\omega_{\rm pl} \gtrsim \omega_{\rm OP}$ must be met in order to activate this strong coupling. On the other hand, under the strong laser excitations, this energy conservation condition loosens up considerably and graphene plasmons with energy $\omega_{\rm pl} < \omega_{\rm OP}$ couple strongly with the hot OP modes. When the plasmon energy is more closer to the interband threshold $2|\varepsilon_F|$, i.e., $\omega_{\rm pl}\approx 0.2$\,eV [see Fig.\,\ref{fig:fig4}(f)], the decay is mostly due to the Landau damping (electron-hole pair interband excitations)\,\cite{bib:jensen91}, which increases with $T_e$ [see Figs.\,\ref{fig:fig3}(e) and \ref{fig:fig3}(f)].

Finally, I would like to emphasize that the presented results are valid for the freely suspended graphene samples. Quantitatively, but not qualitatively, different results are expected for graphene-substrate systems\,\cite{bib:marinko09,bib:gumbs16,bib:iranzo18,bib:vito19,bib:principi18,bib:kim20}. Namely, for graphene-dielectric system one expects only the slight redshift of the plasmon energy\,\cite{bib:marinko09}, while for the graphene-metal contacts the plasmon dispersion obtains linear momentum-dependence and several new features such as higher intensity and lower damping rates\,\cite{bib:principi18,bib:vito19} due to surface-modified electron-electron interactions\,\cite{bib:kim20}. Nevertheless, the present results regarding the non-equilibrium plasmon decay rate are expected to hold also in these graphene-substrate systems.

\section{Conclusion}

Temporal dynamics of non-equilibrium plasmon under intense laser excitation was explored in lightly-doped graphene by means of density functional and density functional perturbation theories. By considering plasmon-phonon coupling, decay rates of graphene Dirac plasmon were studied under both equilibrium (i.e., electron and nuclear degrees of freedom are thermalized) and ultrafast (i.e., electrons and phonons are thermally excited and have disparate energies) conditions. Due to available phase space and energy constraints, equilibrium plasmon with energy $\sim 0.1$\,eV and at ambient temperature is mostly damped due to scatterings with the acoustic phonon modes. Interestingly, the photoinduced non-equilibrium graphene plasmon is, on the other hand, underlain by completely different damping mechanism. Namely, the pump laser pulse increases the population of hot electrons, which in turn transfers the large portion of energy to the strongly coupled optical phonons, creating hot optical-phonon bath. Under such out-of-equilibrium condition phase space for electron scatterings and population of optical phonon modes are increased, as well as the energy constraints are loosened. Consequently, the broadening of the non-equlibrium plasmon is increased immediately after the laser excitation, mostly due to coupling with the hot optical phonons. The corresponding strong plasmon energy renormalization is explained in terms of photoinduced Drude weight increase. For energies $\sim 0.2$\,eV damping pathways of non-equilibrium plasmon are again different and are due to photoinduced interband excitations (Landau damping). Present study might also help elucidate energy-transfer mechanisms under optical pumping in novel quasi-two-dimensional materials that support collective plasmon modes, like metallic or doped semiconducting transition metal dichalcogenides\,\cite{bib:andersen13,bib:scholz13,bib:iurov17b,bib:dajornada20}, topological insulators\,\cite{bib:sim15,bib:in18}, borophene\,\cite{bib:huang17}, buckled-honeycomb lattices\,\cite{bib:tabert14,bib:iurov17}, or layered electrides\,\cite{bib:druffel16,bib:wang19}.

\begin{acknowledgments}
The author acknowledges financial support from the Croatian Science Foundation (Grant no. UIP-2019-04-6869) and from the European Regional Development Fund for the ``Center of Excellence for Advanced Materials and Sensing Devices'' (Grant No. KK.01.1.1.01.0001). Computational resources were provided by the DIPC computing center.
\end{acknowledgments}


\bibliography{tdpl}

\end{document}